\pdfoutput=1
\documentclass[11pt]{article}
\usepackage{amsmath,amsfonts,amssymb,amsthm}
\usepackage[utf8x]{inputenc}
\usepackage[margin=1in]{geometry}
\usepackage{graphicx}
\usepackage{hyperref}
\usepackage{authblk}
\usepackage{times}

\begin{document}

\title{Partitions of Correlated \textit{N}-Qubit Systems}

\author{Simon J. D Phoenix\footnote{
 Department of Physics, Khalifa University, P. O. Box 127788, Abu Dhabi, United Arab Emirates}, Faisal Shah Khan and 
 Berihu Teklu \footnote{
Department of Applied Mathematics and Sciences and Center for Cyber-Physical Systems (C2PS), 
Khalifa University, P. O. Box 127788, Abu Dhabi, United Arab Emirates.}}

\date{\today}
\maketitle
\begin{abstract}

The production and manipulation of quantum correlation protocols will play a central role where 
the quantum nature of the correlation can be used as a resource to yield properties unachievable 
within a classical framework is a very active and important area of research. In this work, we 
provide a description of a measure of correlation strength between quantum systems, 
especially for multipartite quantum systems.

\end{abstract}


\section{Introduction}
In quantum information science, the non-classical correlations in entanglement
play a fundamental role in many protocols. For example, many different topics 
such as phase transition, magnetism, and Bell's inequalities can be analysed by 
the use of correlations. It is well known that the correlation between a system and 
its environment is also a key ingredient in the description of open quantum systems. 
One of the fundamental challenges of measuring correlations in quantum and 
classical systems is that the state of the system can not be directly observed. To 
obtain information one needs to reconstruct the state of the system. This observation 
is complicated in quantum systems and the process of tomography is required. There 
are a variety of ways to quantify quantum correlations~\cite{Adesso16,Horodecki13,Modi12}, 
including the method to quantify spatial correlations in general quantum 
dynamics~\cite{Postler,Angel}, in noisy quantum channels \cite{trapani,berihu,hamza}, 
computation \cite{Galindo} and quantum Schur partitions \cite{Miguel}. 
\par
It is an important task to distinguish and quantify classical and quantum correlations in 
a quantum system. One such example with fundamental informational meaning
 is the quantum mutual information~\cite{Nan,Barnett,Barnett91, Adami,Vedral,Horodecki}
\begin{equation}
I\left( {\rho}\right)  =S\left(  {\rho}_{\alpha}\right)  +S\left(
{\rho}_{\beta}\right)-S\left(  {\rho}_{\alpha\beta}\right)
\end{equation}
where ${\rho}_{\alpha(\beta)}=\text{Tr}_{\beta(\alpha)}[{\rho}_{\alpha\beta}]$ is 
the the reduced density matrix on the subsystem $A(B)$ and $S(\rho)=-\text{Tr}[\rho\log\rho]$ 
denotes the von Neumann entropy. In characterizing correlation one the key questions is to 
ascertain to what extent it can be split into classical and quantum correlations~\cite{Nan}. 
Two well-known properties which depend upon the quantum nature of the correlation 
are quantum entanglement and discord. Entanglement can be used as a resource for 
achieving for many nonlocal information processing tasks~\cite{Ekert,Bennett13} that 
cannot be created between spatially separated subsystems using local operations and 
classical communication (LOCC)~\cite{Werner}. 
However, in many other tasks, this measure is known to play no or very minor 
role~\cite{Knill,Braunstein99,Lanyon}, an alternative measure for quantum correlations 
called quantum discord~\cite{Henderson,Ollivier,Modi12,Misra,Maziero,Maziero10,Auccaise,Auccaise11,Maziero12}, is used as a necessary resource~\cite{Horodecki05,Datta,Madkok}, although there is ongoing controversies 
in the notions of classicality~\cite{Paris}.  
\par

In this work we examine the information approach to correlation strength for multipartite 
quantum systems, initially proposed in~\cite{Barnett93}, that is simply the information content 
of the correlation that a given system possesses. The quantum mutual information is the only 
sensible measure of correlation strength. The index of correlation for two systems is just the 
quantum generalization of the classical mutual information and, for pure states, is equal to 
twice the entropy of entanglement. 

\section{Properties of Correlation}
\label{sec2}

One of the most important and interesting features of a quantum mechanical
description of the world is the existence of correlations that cannot be
explained within a wholly classical theory. There are several useful measures
of the correlation between two quantum systems which all point to this divide
between quantum and classical. Describing the correlation between three or more
systems is, however, more problematical. The interplay between the various
pairwise correlations cannot be easily categorized for a multi-component
quantum system. These multi-component pairwise correlations have interesting
features in the quantum domain, such as the property of monogamy. Here we seek
to define a single measure for the total correlation of a multi-component
system. We have already suggested such a measure, but does this quantity
accord with the general properties we might require of a measure of the total correlation?

Let us consider the general features we might require of any measure of the
`amount' of correlation that a given system possesses. Firstly, we would
require that any such measure is basis-independent. It should not depend on
any particular choice of observable basis but must be a property of the state.
It should return a single number, greater than or equal to zero, that gives a
measure of the amount of correlation that a given state posseses. Secondly we
would require this measure to be additive. If our total system, comprised of
$A$ and $B$, is such that there is no correlation between these component
sub-systems, then the total correlation must simply be the sum of the
correlation \textit{within} $A$ and the correlation \textit{within} $B$. If we
were given $A$ without reference to $B$ then we could, in principle, determine
the strength of correlation of the constituent parts of $A$. Our assessment of
this strength of correlation within $A$ will not change if we subsequently
learn that $A$ is, in fact, correlated with $B$. Our first two general
required properties for our measure of correlation, which we label $I$, are then

\begin{description}
\item[(i)] $I\left(  \hat{\rho}\right)  \geq0$

\item[(ii)] $I\left(  \hat{\rho}\otimes\hat{\sigma}\right)  =I\left(
\hat{\rho}\right)  +I\left(  \hat{\sigma}\right)  $
\end{description}

Property (ii) is suggestive of a logarithmic measure and also specifies what
is meant by identifying $I$ as a measure of `internal' correlation. The
requirement that the measure returns a positive number implies that it is a
trace of some function of the density operator. Hence we seek a measure of the
form%
\begin{equation} \label{eq:intcorr}
I\left(  \hat{\rho}\right)  =\text{Tr}\left[  f\left(  \hat{\rho}\right)
\right]  \geq0
\end{equation}
If the eigenvalues of $\hat{\rho}$ and $\hat{\sigma}$ are $\rho_{j}$ and
$\sigma_{k}$ then property (ii) gives us the requirement that for states of
the form $\hat{\rho}\otimes\hat{\sigma}$ we have%
\begin{equation}\label{eq:add}
\sum\limits_{j}\sum\limits_{k}f\left(  \rho_{j}\sigma_{k}\right)  =\sum
_{j}f\left(  \rho_{j}\right)  +\sum_{k}f\left(  \sigma_{k}\right)
\end{equation}
A function satisfying Eq. (\ref{eq:add}) will be some linear combination of the
entropy functionals $S\left(  \hat{\Omega}\right)  =-$Tr$\left[  \hat{\Omega
}\ln\hat{\Omega}\right]  $ for the components $\hat{\Omega}\in\left\{
\hat{\rho},\hat{\sigma},\hat{\rho}\otimes\hat{\sigma}\right\}  $. This is, of
course, nothing more than a generalization of the derivation of the
information function in communication theory where a similar condition to
property (ii) applied to single events yields the Cauchy functional form for
the information function.

Let us suppose we choose a particular partition of a system into just two 
components which we label $\alpha$ and $\beta$, then property (ii) implies
that the total correlation is of the form%

\begin{equation}
I\left(  \hat{\rho}\right)  =I\left(  \hat{\rho}_{\alpha}\right)  +I\left(
\hat{\rho}_{\beta}\right)  +E\left(  \hat{\rho}_{\alpha},\hat{\rho}_{\beta
}\right)
\end{equation}
where $E\left(  \hat{\rho}_{\alpha},\hat{\rho}_{\beta}\right)  $ is a measure
of the correlation \textit{between} the chosen $\alpha$ and $\beta$ partition,
and which must also be a linear combination of entropy functionals. $E\left(
\hat{\rho}_{\alpha},\hat{\rho}_{\beta}\right)$ can then be interpreted as an
`external' correlation between our chosen partitions. The external correlation
is thus%
\begin{equation}
E\left(\hat{\rho}_{\alpha},\hat{\rho}_{\beta}\right)  =I\left(  \hat{\rho
}\right)  -I\left(  \hat{\rho}_{\alpha}\otimes\hat{\rho}_{\beta}\right)
\end{equation}
which has the natural interpretation as just the difference in the correlation
when our system is viewed as a complete system and when the component parts
are viewed without reference to one another. The requirement that the measure
be a linear combination of the entropies implies that for a two-component system
we must have $E\left(  \hat{\rho}_{\alpha},\hat{\rho}_{\beta}\right)  =\lambda
S\left(  \hat{\rho}_{\alpha}\right)  +\mu S\left(  \hat{\rho}_{\beta}\right)
+\nu S\left(  \hat{\rho}\right)  $. Applying this to a system in which the
$\alpha$ and $\beta$ components represent identical systems with no internal
correlation yields our measure of external correlation as%

\begin{equation}
E\left(  \hat{\rho}_{\alpha},\hat{\rho}_{\beta}\right)  =S\left(  \hat{\rho
}_{\alpha}\right)  +S\left(  \hat{\rho}_{\beta}\right)  -S\left(  \hat{\rho
}\right)
\end{equation}
which has been termed the `index of correlation' and for pure states of the
total system it is twice the entropy of entanglement. For classical systems this 
measure has the natural, and fundamental, interpretation that it is the amount 
of information contained in the external correlation between components 
$\alpha$ and $\beta$. Where these components have no internal degree of 
correlation then $E\left(\hat{\rho}_{\alpha},\hat{\rho}_{\beta}\right)$ is just the internal 
correlation of the total system.

Let us now consider a three-component system such that $I$ for any single
component is zero. We label these components as $A,B$ and $C$. Clearly we can
notionally split this into just two systems so that we have either $\left[
AB,C\right]  ,\left[  BC,A\right]  $ or $\left[  AC,B\right]$. The total
correlation $I\left[ ABC\right]  $ cannot depend on our choice of notional
cut so that we have a further property

\begin{description}
\item[(iii)] the correlation of a multi-component system is invariant of how
we choose to partition the system. The correlation \textit{between} the
components of our chosen partition, however, can vary according to our choice
of partition. Internal and external correlations of the component parts of any
partition are thus \textit{relative} to a particular choice of partitioning.
\end{description}

The amount of correlation in our three component system (where each component has
no degree of internal correlation) can be expressed as a linear combination of
the entropies so that $I\left[  ABC\right]$ must be of the form%
\begin{align}
I\left[  ABC\right]   & =\lambda S\left(  \hat{\rho}_{A}\right)  +\mu S\left(
\hat{\rho}_{B}\right)  +\nu S\left(  \hat{\rho}_{C}\right) \nonumber\\
& +xS\left(  \hat{\rho}_{AB}\right)  +yS\left(  \hat{\rho}_{AC}\right)
+zS\left(  \hat{\rho}_{BC}\right)  +gS\left(  \hat{\rho}\right) \nonumber\\
&
\end{align}
As for the two-component system, consideration of three identical uncorrelated
systems fixes this quantity to be%
\begin{equation}
I\left[  ABC\right]  =S\left(  \hat{\rho}_{A}\right)  +S\left(  \hat{\rho}%
_{B}\right)  +S\left(  \hat{\rho}_{C}\right)  -S\left(  \hat{\rho}\right)
\end{equation}
with the obvious generalization to an $n$-component system of%
\begin{equation}\label{eq:gen}
I\left[  12\ldots n\right]  =\sum\limits_{k=1}^{n}S\left(  \hat{\rho}%
_{k}\right)  -S\left(  \hat{\rho}\right)
\end{equation}
This has the natural and appealing interpretation that the correlation
contained in an $n$-component system is the difference in the information
obtained when looking at joint properties and the information obtained when
looking at each system in isolation. It is straightforward to show that
Eq. (\ref{eq:gen}) satisfies the three natural requirements above for a measure
of correlation.
 \section{Correlated Quantum Systems}
\label{sec3}

Consider $N$ quantum systems (assumed to possess no degree of internal
correlation) which we label by the index $1\leq k\leq N$, then the total
information content of the correlation that exists between the systems is
given by \cite{herbut, maziero}%
\begin{equation}\label{eq:total}
I\left(  N\right)  =\sum\limits_{k=1}^{N}S_{k}-S
\end{equation}
where $S_{k}$ are the individual sub-system entropies and $S$ is the entropy
of the complete system. The word total here simply refers to the totality of information 
that is carried by the correlation between all the subsystems. The interpretation of 
Eq. (\ref{eq:total}) is that the information contained in the correlation is the difference 
in the information content when considering properties of the sub-systems 
alone and when considering the complete system. This difference in information is 
clearly 'residing' in the correlation and manifests in correlated joint properties. 
Actually using, or accessing, this correlational information directly is another 
matter entirely. The arguments above show that there is information in the correlation, 
but not how to access that or to use it for coding, for example. The entropies are defined 
in the usual way through the von Neumman entropy as%
\begin{align}
S  & =-\text{Tr}\left\{  \rho\ln\rho\right\} \nonumber\\
& \nonumber\\
S_{k}  & =-\text{Tr}_{k}\left\{  \rho_{k}\ln\rho_{k}\right\} \nonumber\\
& \nonumber\\
\text{ with \ \ }\rho_{k}  & =\text{Tr}_{\text{all }j\neq k}~\left\{
\rho\right\}
\end{align}
The index of correlation (1) has different upper bounds for quantum and
classical systems. Consider a system of $N$ quantum sub-systems where, $wlog$,
we have that%

\begin{equation}
S_{1}\geq S_{2}\geq\ldots\geq S_{N}%
\end{equation}
If these were classical systems we would have that%

\begin{equation}
\sup\left\{  S_{1},S_{2},\ldots,S_{N}\right\}  \leq S\leq S_{1}+S_{2}%
+\ldots+S_{N}%
\end{equation}
Hence, classically the upper bound on the total information content of the
correlation is given by%

\begin{equation}
I_{C}\left(  N\right)  \leq%
{\displaystyle\sum\limits_{j=2}^{N}}
S_{j}%
\end{equation}
For quantum systems, however, the total entropy can be equal to zero so that
an upper bound for the information content of the correlation is%

\begin{equation}
I_{Q}\left(  N\right)  \leq%
{\displaystyle\sum\limits_{j=1}^{N}}
S_{j}%
\end{equation}
The difference between the two quantities is bounded by%

\begin{equation}
I_{Q}\left(  N\right)  -I_{C}\left(  N\right)  \leq S_{1}%
\end{equation}
or by dropping the ordering convention of (3) more generally as%
\begin{equation}
I_{Q}\left(  N\right)  -I_{C}\left(  N\right)  \leq\sup\left\{  S_{1}%
,S_{2},\ldots,S_{N}\right\}
\end{equation}
The correlation in the classical and quantum cases is bounded by%
\begin{align}
0  & \leq I_{C}\left(  N\right)  \leq\inf\left\{  S_{1},S_{2},\ldots
,S_{N}\right\} \nonumber\\
0  & \leq I_{Q}\left(  N\right)  \leq2\inf\left\{  S_{1},S_{2},\ldots
,S_{N}\right\}
\end{align}
where by `classical' and `quantum' here we mean that given a system of
correlated objects a classical description of those objects necessarily
satisfies the entropy bound (4), but a quantum description has a lower bound
of zero for its entropy because of the potential for the system being in a
pure state. If we use a tilde to denote the maximum possible entropy
attainable by a system (or sub-system) then we can describe the region $0\leq
I\left(  N\right)  \leq\inf\left\{  \tilde{S}_{1},\tilde{S}_{2},\ldots
,\tilde{S}_{N}\right\}  $ as a classical region, in that classical states
exist for the system under consideration with this strength of correlation. A
system having a strength of correlation in this region is not
\textit{necessarily} classical. There are correlated quantum states with a
strength of correlation in this region which will, nevertheless, display
non-classical correlation properties. However, given such a quantum state in
this region, it is always possible to find at least one classical state of the
system which possesses the same correlation strength. The region $\inf\left\{
\tilde{S}_{1},\tilde{S}_{2},\ldots,\tilde{S}_{N}\right\}  \leq I\left(
N\right)  \leq2\inf\left\{  \tilde{S}_{1},\tilde{S}_{2},\ldots,\tilde{S}%
_{N}\right\}  $ is a strength of correlation that \textit{cannot} be attained
by \textit{any} classical state of the system. Correlation strengths, as
measured by (1), which have a value in this region can only be attained by
quantum states of the system. It is with these remarks in mind that we
describe these regions of correlation strength as classical and quantum,
respectively. Although we have not been able to prove this in general, it
seems reasonable to argue that correlation strengths in the quantum region
will lead to observable consequences in the correlation properties that are
non-classical. One example of such an observable consequence might be a
violation of a suitable Bell-type inequality.

\subsection{Partitioning of the Systems}
\unskip

We can assign an arbitrary \textit{single} partition on the system so that the
complete system is now split into 2 components such that component $\alpha$
contains $n$ sub-systems and component $\beta$ contains $m$ sub-systems where
$n+m=N$. The index of correlation can now be written as%
\begin{equation}
I\left(  N\right)  =I^{int}\left(  \alpha\right)  +I^{int}\left(
\beta\right)  +I^{ext}\left(  \alpha,\beta\right)
\end{equation}
where $I^{int}\left(  \alpha\right)  $ gives the `internal' correlation (as
measured by the index of correlation (1)) for the component system comprising
$n$ sub-systems, and $I^{ext}\left(  \alpha,\beta\right)  $ gives the
`external' correlation between the two component parts of the partition where
we think each of component being treated as a single entity. Of course, such a
partition is merely notional unless we perform some physical action to
separate the components $\alpha$ and $\beta$. The overall correlation
$I\left(  N\right)  $ is independent of our notional partition, but the
`internal' and `external' components are not. The notions of internal and
external are \textit{relative} to a chosen partition.

A simple way of making the partition physically real is to physically separate
our quantum system into 2 distinct `boxes'. We might then give one box to
Alice (say) and one box to Bob. We can now ask whether (given an ensemble of
such boxes) there are measurements Alice (or Bob) can do to reveal evidence of
correlation amongst the sub-systems in their respective boxes. What we have
termed `internal' correlation is simply the information content of the
correlation in each box considered separately. 
\begin{figure}
\resizebox{0.85\columnwidth}{!}{
\includegraphics{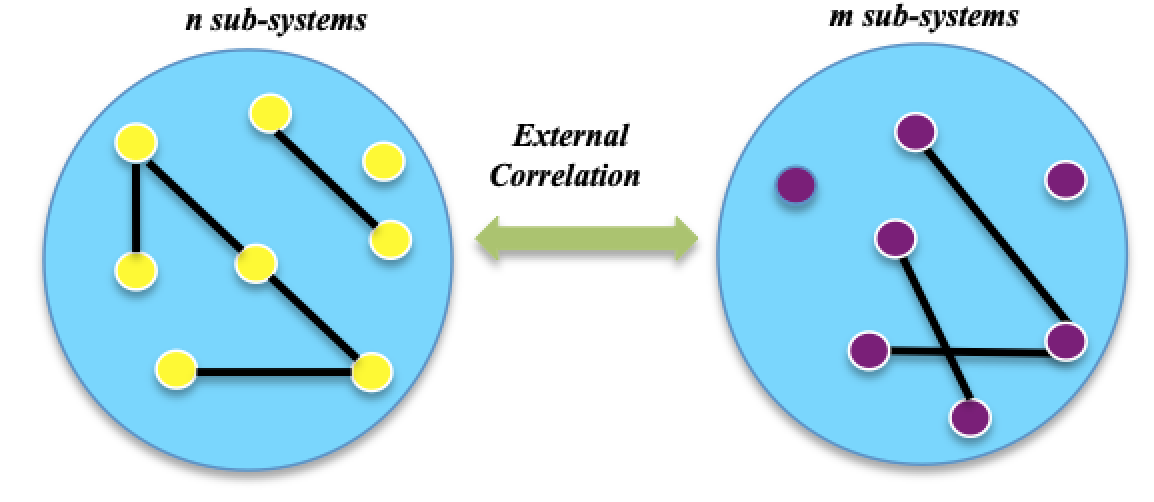} }
\caption{Schematic representation of the subsystems $n$ and $m$.
}
\label{fig1}
\end{figure}  
The index of correlation, whether applied to a complete $n+m$ system or to an
individual component of the partition, returns a single number. If we fix the
overall total correlation $I\left(  N\right)  $ to be some value, then this
does not specify the state uniquely and there are many possible states of the
complete system that will yield this value for the index of correlation. So if
we have two density operators $\rho\left(  1\right)  $ and $\rho\left(
2\right)  $ that yield the same value for total index of correlation we have%
\begin{equation}
I_{1}^{int}\left(  \alpha\right)  +I_{1}^{int}\left(  \beta\right)
+I_{1}^{ext}\left(  \alpha,\beta\right)  =I_{2}^{int}\left(  \alpha\right)
+I_{2}^{int}\left(  \beta\right)  +I_{2}^{ext}\left(  \alpha,\beta\right)
\end{equation}
or equivalently%
\begin{equation}
\left[  I_{1}^{int}\left(  \alpha\right)  +I_{1}^{int}\left(  \beta\right)
\right]  -\left[  I_{2}^{int}\left(  \alpha\right)  +I_{2}^{int}\left(
\beta\right)  \right]  =I_{2}^{ext}\left(  \alpha,\beta\right)  -I_{1}%
^{ext}\left(  \alpha,\beta\right)
\end{equation}
Thus, for the same value of $I\left(  N\right)  $, increasing the degree of
external correlation between the components of a given partition reduces the
combined degree of internal correlation and vice versa. Indeed, if we consider
a partition such that there is no external correlation between the 2
components then the density operator for the total system is separable and can
be written in the form $\rho=\rho_{\alpha}\otimes\rho_{\beta}$.

\subsection{Maximising the Total Correlation}

Let us now consider states of the complete system such that the index of
correlation for the complete system, $I\left(  N\right)  $, is maximised. This
occurs when the total system is in a pure state. If we consider the
sub-systems, $1,2,\ldots,N$ to be the same physical objects (thus $N$
electrons, or $N$ two-level atoms, for example) then the maximum value that
$I\left(  N\right)  $ can take is just $I^{\max}\left(  N\right)  =NS^{\max}$
where $S^{\max}$ is the maximum entropy that a single sub-system can attain.
For pure states of the complete system (whether maximally correlated or not)
when we partition the space into just 2 components, the entropies of those
components are equal. This follows from the Araki-Lieb triangle inequality which 
states that, for any two quantum systems $A$ and $B$, we have%
\begin{equation}
\left\vert S_{A}-S_{B}\right\vert \leq S\leq S_{A}+S_{B}%
\end{equation}
For pure states of a system comprising of $N$ quantum systems, split into just
2 partitions of $n$ and $m$ objects we therefore have that%
\begin{align}
0  & \leq I^{int}\left(  \alpha\right)  +I^{int}\left(  \beta\right)  \leq
NS^{\max}\nonumber\\
0  & \leq I^{ext}\left(  \alpha,\beta\right)  \leq NS^{\max}%
\end{align}
where if we maximise $I^{int}\left(  \alpha\right)  +I^{int}\left(
\beta\right)  $ we minimise $I^{ext}\left(  \alpha,\beta\right)  $, and vice versa.

In general, therefore, where the sub-systems are the same physical entities,
in order to \textit{globally} maximise the total correlation we require the
following 2 conditions to be met

\begin{description}
\item[(\textit{i})] the total system must be in a pure state

\item[(\textit{ii})] each individual sub-system must be in a state of maximum
entropy so that the global maximum correlation strength is given by%
\[
I^{\max}\left(  N\right)  =\sum\limits_{k=1}^{N}\tilde{S}_{k}=NS^{\max}%
\]
\bigskip
\end{description}

In what follows we shall restrict ourselves to \textit{pure states} of the
total $N$-component system. With this condition we can view the total state as
a purification of the $n$-component partition (or the $m$-component
partition). As remarked above, the entropies of these two component partitions
are equal if the total system is in a pure state.

\section{Correlated Systems of $4$ Qubits}
\label{sec4}

To begin with we shall consider $N=4$ so that we have just 4 qubits. This
example is sufficient to illustrate many of the properties of the correlation
between larger systems of qubits. We shall consider states of these 4 qubits
that maximise $I\left(  N\right)  $ which, for 4 qubits, takes the value
$4\ln2$. Clearly, these states have to be pure states of 4 qubits. For 4
qubits there are only 2 possible $\left(  n,m\right)  $ partitions which are
$\left(  1,3\right)  $ and $\left(  2,2\right)  $ and we consider the
partitions $\left(  3,1\right)  $ and $\left(  1,3\right)  $ to be physically
equivalent, by symmetry. Labelling the individual qubits as $a,b,c,d,\ldots$
it is clear from above that for a total state of $N$ qubits to maximise the
information content of the correlations we must have that $S_{a}=S_{b}%
=S_{c}=\ldots=\ln2$. Equivalently, we require that the reduced density
operator for any single qubit to be of the form $\left(  \left\vert
0\right\rangle \left\langle 0\right\vert +\left\vert 1\right\rangle
\left\langle 1\right\vert \right)  /2$.

\subsection{Minimising $I^{ext}\left(  \alpha,\beta\right)  $}

A state that (globally) maximises \textit{both} $I\left(  N\right)  $
\textit{and} $I^{int}\left(  \alpha\right)  +I^{int}\left(  \beta\right)  $ is
not possible for the $\left(  1,3\right)  $ partition of 4 qubits. Such a
state is, however, possible for the $\left(  2,2\right)  $ partition and an
example of such a state is given by%
\begin{equation}
\left\vert \psi\right\rangle _{S}=\frac{1}{\sqrt{2}}\left(  \left\vert
00\right\rangle _{ab}+\left\vert 11\right\rangle _{ab}\right)  \otimes\frac
{1}{\sqrt{2}}\left(  \left\vert 00\right\rangle _{cd}+\left\vert
11\right\rangle _{cd}\right)
\end{equation}
where the states to the left of the tensor product in this expression refer to
one component partition (containing the qubits labelled $a$ and $b$) and those
to the right the other. This state minimises the external correlation between
the \textit{chosen} component partitions and is, thus, separable being a
tensor product of two maximally correlated states of 2 qubits. It satisfies
the constraint that the correlation of the total 4 qubit system is maximised.

It is important to note that, in general for a given $\left\vert
\psi\right\rangle $ which maximises $I\left(  N\right)  $, the external
correlation between partitions is invariant under permutations of particles
\textit{within} each component partition, but not invariant to permutations of
particles \textit{between} the partitions. As we shall see the GHZ state of
$N$ qubits is a highly symmetric state in that the external correlation is
invariant under permutations between partitions, and also invariant of the
number of particles within each partition. The 4 qubit state (10) gives a nice
example of the effect of permutation between the partitions. If we swap one
particle from $\alpha$ with a particle from $\beta$ the resultant density
operator is not separable in terms of the reduced density operators of the new
$\alpha$ and $\beta$ components.
\subsection{Maximising $I^{ext}\left(  \alpha,\beta\right)  $}

An example of a state that (globally) maximises \textit{both} $I\left(
N\right)  $ \textit{and} $I^{ext}\left(  \alpha,\beta\right)  $ for the
$\left(  2,2\right)  $ partition is given by%
\begin{equation}
\left\vert \psi\right\rangle _{UE}=\frac{1}{2}\left(  \left\vert
00,00\right\rangle +\left\vert 01,01\right\rangle +\left\vert
10,10\right\rangle +\left\vert 11,11\right\rangle \right)
\end{equation}
where the comma in the state label distinguishes between the component
partitions. If we label the individual qubits as $a,b,c,d$ and the two
component partitions as $\alpha$ and $\beta$, so that the component partition
$\alpha$ is comprised of the qubits $a$ and $b$, then it is easy to see that
state (15) yields the reduced density operators%
\begin{align}
\rho_{\alpha\left(  \beta\right)  }  & =\frac{1}{2}\left(  \left\vert
00\right\rangle \left\langle 00\right\vert +\left\vert 01\right\rangle
\left\langle 01\right\vert +\left\vert 10\right\rangle \left\langle
10\right\vert +\left\vert 11\right\rangle \left\langle 11\right\vert \right)
\nonumber\\
& \nonumber\\
\rho_{a\left(  b\right)  }  & =\frac{1}{2}\left(  \left\vert 0\right\rangle
\left\langle 0\right\vert +\left\vert 1\right\rangle \left\langle 1\right\vert
\right)
\end{align}
It is clear that, \textit{given just one of the 2 component partitions}, no
evidence of correlation can be detected by any measurement. This is a general
property of any state of $N$ systems that (globally) maximises the external
correlation of a partition (for $N$ even this simply implies a partition such
that $n=m=N/2$). This property is often expressed as a condition for `maximal'
entanglement in that this maximally `mixes' the component systems.

\subsection{Intermediate States}

Now that we've seen examples of extremal states (maximally correlated states
of 4 qubits that maximise the internal correlation, or the external
correlation) we consider a state intermediate between these extremes. An
obvious candidate is the GHZ state of 4 qubits, an example of which is given by%

\begin{equation}
\left\vert \psi\right\rangle _{GHZ}=\frac{1}{\sqrt{2}}\left(  \left\vert
0000\right\rangle +\left\vert 1111\right\rangle \right)
\end{equation}
For this state the reduced density operators in the $\left(  2,2\right)  $
partition are given by
\begin{align}
\rho_{\alpha(\beta)}  & =\frac{1}{2}\left(  \left\vert 00\right\rangle
\left\langle 00\right\vert +\left\vert 11\right\rangle \left\langle
11\right\vert \right) \nonumber\\
& \nonumber\\
\rho_{a(b)}  & =\frac{1}{2}\left(  \left\vert 0\right\rangle \left\langle
0\right\vert +\left\vert 1\right\rangle \left\langle 1\right\vert \right)
\end{align}
The internal correlation of one component is, therefore, $I^{int}\left(
\alpha\right)  =\ln2$ and the external correlation between the components is
$I^{ext}\left(  \alpha,\beta\right)  =2\ln2$. We can see that as we vary the
state from $\left\vert \psi\right\rangle _{GHZ}$ to $\left\vert \psi
\right\rangle _{S}$, \textit{whilst keeping} $I\left(  N\right)  $
\textit{fixed at its maximum value of} $4\ln2$, we have that
\begin{equation}
\ln2\leq I^{int}\left(  \alpha\right)  \leq2\ln2
\end{equation}
  
\begin{figure}
\resizebox{0.85\columnwidth}{!}{
\includegraphics{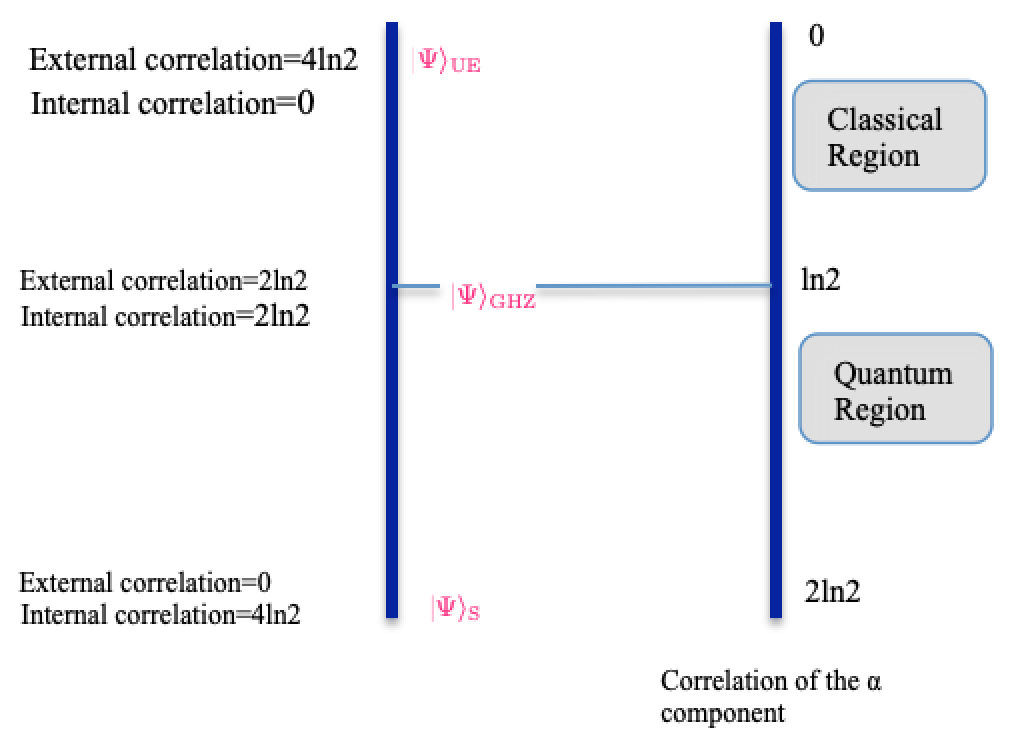} }
\caption{Schematic representation of the system and correlations considered.
}
\label{fig2}
\end{figure}  

but is this is precisely the non-classical region for the correlation of 2
qubits as expressed by (8). If we are given a 2-qubit component of a maximally
correlated state of 4 qubits, then the GHZ state $\left\vert \psi\right\rangle
_{GHZ}$ acts as a boundary between classical and quantum regions of the
correlations for that component. This is summarized in the figure below.
The states on the left hand side of this figure represent all possible
purifications of 2 qubits, \textit{for the chosen partition}, obtained by
adding 2 extra qubits, that yield a maximally correlated pure state. In order
to satisfy the requirement that this purification gives a maximally correlated
state of 4 qubits we must have that%
\begin{equation}
\rho_{a(b)}=\frac{1}{2}\left(  \left\vert 0\right\rangle \left\langle
0\right\vert +\left\vert 1\right\rangle \left\langle 1\right\vert \right)
\end{equation}

In the quantum regime for the correlations of 2 qubits we can see that a
purification can be achieved by adding a \textit{single extra qubit}. States
of 2 qubits in this region can be purified by adding a single extra qubit
giving a pure state of 3 qubits of the form%

\begin{equation}
\left\vert \xi\right\rangle =\mu\left\vert 0\right\rangle \otimes\left\vert
\pi_{1}\right\rangle +\nu\left\vert 1\right\rangle \otimes\left\vert \pi
_{2}\right\rangle
\end{equation}
which (except for the initial GHZ state at the boundary between the regions)
is not a maximally correlated state of 3 qubits. Thus the states of 2 qubits,
such that each individual qubit is in a maximally mixed state, which display
non-classical correlation properties are \textit{precisely those that can be
purified by the addition of a single qubit}. In the classical regime for the
correlations of 2 qubits we need at least 2 qubits to effect a purification,
for the \textit{chosen} partition. It seems reasonable to suppose that in this
quantum region (where the 2 qubits originate from a maximally-correlated state
of 4 qubits) we will obtain a violation of a suitably-chosen Bell inequality
for the 2 qubits.

It should be noted that where the initial state is placed on the left hand
side of figure 2 is dependent on which 2 qubits are chosen for the partition.
If we begin with an initial state of the 4 qubits given by equation (15) then
the above discussion assumes that qubits $a$ and $b$ were chosen for one
partition and qubits $c$ and $d$ for the other. If, however, we choose the
partitions $\left(  a,c\right)  $ and $\left(  b,d\right)  $ then, for these
partitions, the external correlation is maximised whilst the internal
correlation is zero. 
\section{Equal Partitions of Correlated Systems of $N$ Qubits}
\label{sec5}
We now consider the effect of a single partition, as above, on systems of $N$
qubits. As we are going to consider a partitioning into 2 collections of equal
numbers of qubits we shall take $N$ even so that $N=2n$ and label the qubits
as $a_{1},a_{2},\ldots a_{2n}$. We consider pure states of the $2n$ qubits
such that the entropy of any single qubit is $S\left(  a_{k}\right)  =\ln2$
(that is, 1 bit) and that $I\left(  N\right)  =2n\ln2$ (the pure state
maximises the information content of the correlation for the entire system of
$2n$ qubits). With these conditions we can write the internal and external
correlations as%
\begin{align}
I^{ext}\left(  \alpha,\beta\right)   & =2S\left(  \alpha\right) \nonumber\\
I^{int}\left(  \alpha\right)   & =n\ln2-S\left(  \alpha\right) \nonumber\\
I^{int}\left(  \beta\right)   & =n\ln2-S\left(  \alpha\right)
\end{align}
As before, we can consider an `extremal' state that maximises the external
correlation which is%
\begin{equation}
\left\vert \psi\right\rangle _{UE}=\frac{1}{\sqrt{2^{n}}}\sum\limits_{s}%
\left\vert s,s\right\rangle
\end{equation}
where $s$ is an index, written in binary, such that $0\leq s\leq2^{n}-1$ and
the comma distinguishes the $\alpha$ and $\beta$ components. The index $s$ is
just a bit string that ranges over the $2^{n}-1$ levels of the $\alpha$ and
$\beta$ components. The internal correlations of the $\alpha$ and $\beta$
components.$I^{int}\left(  \alpha\right)  =I^{int}\left(  \beta\right)  =0$
for this state. As an example of the opposite extreme we can consider the
state%
\begin{equation}
\left\vert \psi\right\rangle _{S}=\frac{1}{\sqrt{2}}\left(  \left\vert
0\right\rangle ^{n}+\left\vert 1\right\rangle ^{n}\right)  \otimes\frac
{1}{\sqrt{2}}\left(  \left\vert 0\right\rangle ^{n}+\left\vert 1\right\rangle
^{n}\right)
\end{equation}
which separates the $\alpha$ and $\beta$ components into two (maximally
correlated) pure states. The state $\left\vert \psi\right\rangle _{S}$
maximises the internal correlations of the $\alpha$ and $\beta$ components but
there is no external correlation and $I^{ext}\left(  \alpha,\beta\right)  =0$.
As above we consider a GHZ state of the $2n$ qubits as an example of a state
intermediate between these 2 extremes and this can be written as%
\begin{equation}
\left\vert \psi\right\rangle _{GHZ}=\frac{1}{\sqrt{2}}\left(  \left\vert
0\right\rangle ^{2n}+\left\vert 1\right\rangle ^{2n}\right)
\end{equation}
The properties of internal and external correlations for these states are
summarized in the figure below.

\begin{figure}
\resizebox{0.8\columnwidth}{!}{
\includegraphics{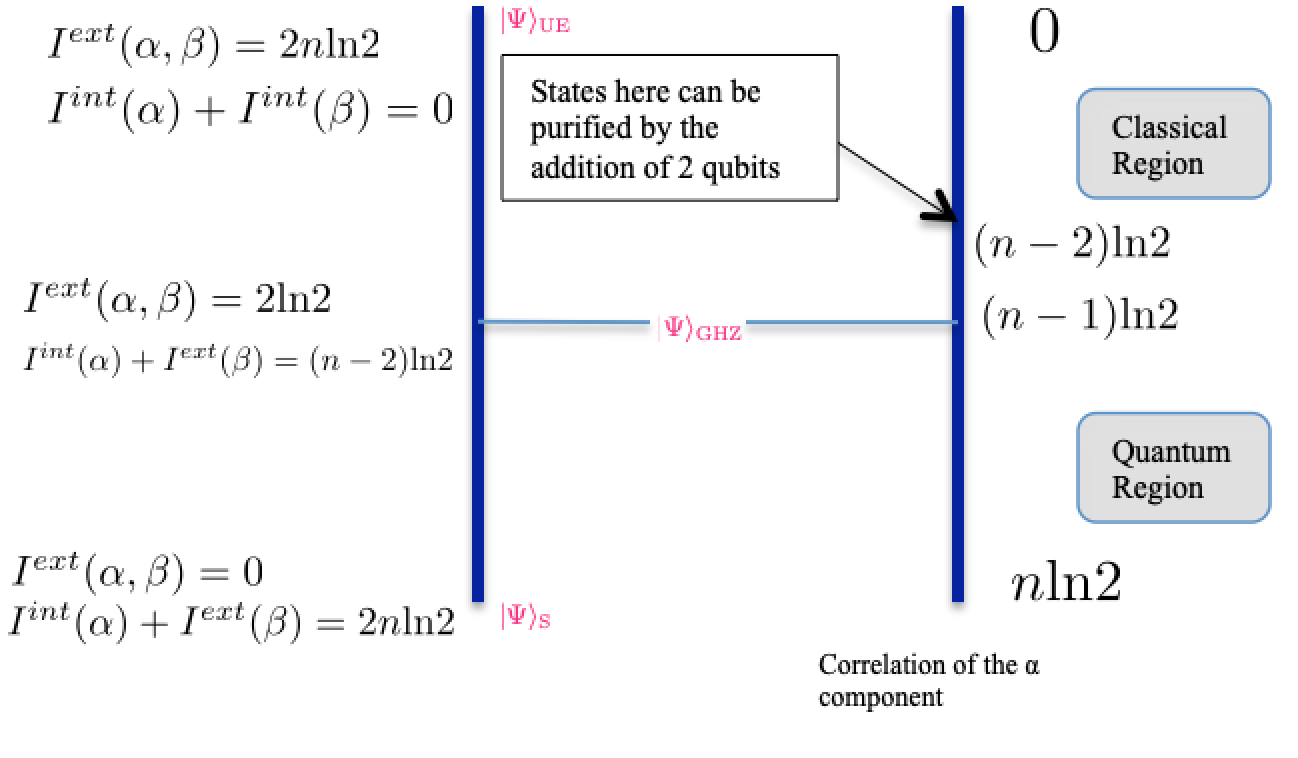} }
\caption{Schematic representation of the system and correlations considered.
}
\label{fig3}
\end{figure}   
The GHZ state again forms a kind of boundary between `classical' and
`quantum'. States of the $\alpha$-partition which yield $I^{int}\left(
\alpha\right)  $ in this quantum region are, necessarily, quantum mechanical
in nature. States which yield $I^{int}\left(  \alpha\right)  $ in the
`classical' region may indeed display non-classical features, but in this
region it is possible to find a \textit{classical} state of the $n$ qubits
with the same correlation strength. States of $n$ qubits, where each
individual qubit has an entropy of $S\left(  a_{k}\right)  =\ln2$, that can be
purified by the addition of a \textit{single} qubit, are in this quantum
region. In the `classical' region we need to add 2, or more, qubits to achieve
a purification. The resultant purifications achieved are not maximally
correlated states of $n+k$ qubits (where $k$ is the number of qubits we need
to add to achieve purification), except for the case of the $n $-qubit
$\alpha$ component that has arisen from the GHZ state of $2n$ qubits.

The GHZ state has a high degree of symmetry in the following sense. If we
start with our $2n$ qubits prepared in the GHZ state and partition into
$\left(  n,m\right)  $ qubits, then for this state the external correlation
$I^{ext}\left(  \alpha,\beta\right)  =2\ln2$, independently of where we make
the cut or the number of qubits in each partition. The external correlation
for the GHZ state is invariant under permutations of the particles in a given
partition, and is invariant of the number of particles in each partition.

\section{Conclusions}
We have studied a measure of correlation strength for multipartite quantum systems, that is 
the information content of the correlation. It is important to emphasize that this is a measure 
of correlation strength only; it does not distinguish the precise nature of the correlation itself. 
However, we have shown that if this correlation strength is above a certain value then that can 
only be achieved by quantum mechanisms. This is defined as a natural extension of the index 
of correlation for bipartite systems. The index of correlation for two systems is just the quantum 
generalization of the classical mutual information and, for pure states, is equal to twice the entropy 
of entanglement. This information-based measure arises as a consequence of imposing certain 
natural conditions on any measure of correlation strength. In this note we have used this measure, 
together with the notion of partitioning, to derive some general properties of the correlation of 
interacting quantum systems. The term partitioning is only notional unless we take 
steps to physically create the partitions,  but it allows us to identify the various entropy 
and correlation invariants of the interaction. Once these invariants have been identified 
it only requires very elementary techniques to establish these general properties.  


\subsection{Acknowledgments}{BT greatly acknowledge supports from the Center for Cyber-Physical Systems (C2PS)-Khalifa University, Abu Dhabi, UAE.}

\end{document}